\RequirePackage{lineno} 
\documentclass[aps,prd,twocolumn,showpacs,superscriptaddress,floatfix,amsmath]{revtex4}

\usepackage{graphicx} 
\usepackage{dcolumn} 
\usepackage{bm} 
\usepackage{epstopdf}
\usepackage{bm}        
\usepackage{amssymb}   
\newcommand{\MET}{\ensuremath{{\slash\kern-.7emE}_{T}}}
\newcommand{\pte}{p_T^{e}}
\newcommand{\mte}{m_T^{e}}
\newcommand{\ptmu}{p_T^{\mu}}
\newcommand{\mtmu}{m_T^{\mu}}
\newcommand{\METmu}{\MET^{\mu}}
\newcommand{\METe}{\MET^{e}}
\newcommand{\METl}{\MET^{\ell}}

\newcommand{\TeV}{\ensuremath{\mathrm{Te\kern -0.1em V}}}
\newcommand{\GeV}{\ensuremath{\mathrm{Ge\kern -0.1em V}}}
\newcommand{\MeV}{\ensuremath{\mathrm{Me\kern -0.1em V}}}
\def\GeVc2{\ensuremath{\mathrm{ Ge\kern -0.1em V }\kern -0.2em /c^2 }}

\newcommand{\MW}{\ensuremath{M_{W}}}
\newcommand{\MZ}{\ensuremath{M_{ Z }}}
\newcommand{\GW}{\ensuremath{\Gamma_{W }}}

\newcommand{\bluealle}{19}

\newcommand{\syspscalecombin}{7}

\newcommand{\sysholecombin}{2}

\newcommand{\sysbkgcombin}{3}
\newcommand{\sysgtwocombin}{5}
\newcommand{\sysqedcombin}{4}
\newcommand{\syspdfcombin}{10}
\newcommand{\statcombin}{12}

\newcounter {subsubsubsection}[subsubsection]

\def\lsim{\mathrel{\rlap{\lower4pt\hbox{\hskip1pt$\sim$}}\raise1pt\hbox{$<$}}}
\newcommand{\metsub}{\ensuremath{{\slash\kern-.5emE}_{T}}}

\begin{document}

\title{Combination of CDF and D0 $\bm{W}$-Boson Mass Measurements}
\affiliation{Institute of Physics, Academia Sinica, Taipei, Taiwan 11529, Republic of China}
\affiliation{Argonne National Laboratory, Argonne, Illinois 60439, USA}
\affiliation{University of Athens, 157 71 Athens, Greece}
\affiliation{Institut de Fisica d'Altes Energies, ICREA, Universitat Autonoma de Barcelona, E-08193, Bellaterra (Barcelona), Spain}
\affiliation{Baylor University, Waco, Texas 76798, USA}
\affiliation{Istituto Nazionale di Fisica Nucleare Bologna, \ensuremath{^{uu}}University of Bologna, I-40127 Bologna, Italy}
\affiliation{University of California, Davis, Davis, California 95616, USA}
\affiliation{University of California, Los Angeles, Los Angeles, California 90024, USA}
\affiliation{Instituto de Fisica de Cantabria, CSIC-University of Cantabria, 39005 Santander, Spain}
\affiliation{Carnegie Mellon University, Pittsburgh, Pennsylvania 15213, USA}
\affiliation{Enrico Fermi Institute, University of Chicago, Chicago, Illinois 60637, USA}
\affiliation{Comenius University, 842 48 Bratislava, Slovakia; Institute of Experimental Physics, 040 01 Kosice, Slovakia}
\affiliation{Joint Institute for Nuclear Research, RU-141980 Dubna, Russia}
\affiliation{Duke University, Durham, North Carolina 27708, USA}
\affiliation{Fermi National Accelerator Laboratory, Batavia, Illinois 60510, USA}
\affiliation{University of Florida, Gainesville, Florida 32611, USA}
\affiliation{Laboratori Nazionali di Frascati, Istituto Nazionale di Fisica Nucleare, I-00044 Frascati, Italy}
\affiliation{University of Geneva, CH-1211 Geneva 4, Switzerland}
\affiliation{Glasgow University, Glasgow G12 8QQ, United Kingdom}
\affiliation{Harvard University, Cambridge, Massachusetts 02138, USA}
\affiliation{Division of High Energy Physics, Department of Physics, University of Helsinki, FIN-00014, Helsinki, Finland; Helsinki Institute of Physics, FIN-00014, Helsinki, Finland}
\affiliation{University of Illinois, Urbana, Illinois 61801, USA}
\affiliation{The Johns Hopkins University, Baltimore, Maryland 21218, USA}
\affiliation{Institut f\"{u}r Experimentelle Kernphysik, Karlsruhe Institute of Technology, D-76131 Karlsruhe, Germany}
\affiliation{Center for High Energy Physics: Kyungpook National University, Daegu 702-701, Korea; Seoul National University, Seoul 151-742, Korea; Sungkyunkwan University, Suwon 440-746, Korea; Korea Institute of Science and Technology Information, Daejeon 305-806, Korea; Chonnam National University, Gwangju 500-757, Korea; Chonbuk National University, Jeonju 561-756, Korea; Ewha Womans University, Seoul, 120-750, Korea}
\affiliation{Ernest Orlando Lawrence Berkeley National Laboratory, Berkeley, California 94720, USA}
\affiliation{University of Liverpool, Liverpool L69 7ZE, United Kingdom}
\affiliation{University College London, London WC1E 6BT, United Kingdom}
\affiliation{Centro de Investigaciones Energeticas Medioambientales y Tecnologicas, E-28040 Madrid, Spain}
\affiliation{Massachusetts Institute of Technology, Cambridge, Massachusetts 02139, USA}
\affiliation{Institute of Particle Physics: McGill University, Montr\'{e}al, Qu\'{e}bec H3A~2T8, Canada; Simon Fraser University, Burnaby, British Columbia V5A~1S6, Canada; University of Toronto, Toronto, Ontario M5S~1A7, Canada; TRIUMF, Vancouver, British Columbia V6T~2A3, Canada}
\affiliation{University of Michigan, Ann Arbor, Michigan 48109, USA}
\affiliation{Michigan State University, East Lansing, Michigan 48824, USA}
\affiliation{Institution for Theoretical and Experimental Physics, ITEP, Moscow 117259, Russia}
\affiliation{University of New Mexico, Albuquerque, New Mexico 87131, USA}
\affiliation{The Ohio State University, Columbus, Ohio 43210, USA}
\affiliation{Okayama University, Okayama 700-8530, Japan}
\affiliation{Osaka City University, Osaka 558-8585, Japan}
\affiliation{University of Oxford, Oxford OX1 3RH, United Kingdom}
\affiliation{Istituto Nazionale di Fisica Nucleare, Sezione di Padova, \ensuremath{^{vv}}University of Padova, I-35131 Padova, Italy}
\affiliation{University of Pennsylvania, Philadelphia, Pennsylvania 19104, USA}
\affiliation{Istituto Nazionale di Fisica Nucleare Pisa, \ensuremath{^{ww}}University of Pisa, \ensuremath{^{xx}}University of Siena, \ensuremath{^{yy}}Scuola Normale Superiore, I-56127 Pisa, Italy, \ensuremath{^{zz}}INFN Pavia, I-27100 Pavia, Italy, \ensuremath{^{aaa}}University of Pavia, I-27100 Pavia, Italy}
\affiliation{University of Pittsburgh, Pittsburgh, Pennsylvania 15260, USA}
\affiliation{Purdue University, West Lafayette, Indiana 47907, USA}
\affiliation{University of Rochester, Rochester, New York 14627, USA}
\affiliation{The Rockefeller University, New York, New York 10065, USA}
\affiliation{Istituto Nazionale di Fisica Nucleare, Sezione di Roma 1, \ensuremath{^{bbb}}Sapienza Universit\`{a} di Roma, I-00185 Roma, Italy}
\affiliation{Mitchell Institute for Fundamental Physics and Astronomy, Texas A\&M University, College Station, Texas 77843, USA}
\affiliation{Istituto Nazionale di Fisica Nucleare Trieste, \ensuremath{^{ccc}}Gruppo Collegato di Udine, \ensuremath{^{ddd}}University of Udine, I-33100 Udine, Italy, \ensuremath{^{eee}}University of Trieste, I-34127 Trieste, Italy}
\affiliation{University of Tsukuba, Tsukuba, Ibaraki 305, Japan}
\affiliation{Tufts University, Medford, Massachusetts 02155, USA}
\affiliation{University of Virginia, Charlottesville, Virginia 22906, USA}
\affiliation{Waseda University, Tokyo 169, Japan}
\affiliation{Wayne State University, Detroit, Michigan 48201, USA}
\affiliation{University of Wisconsin, Madison, Wisconsin 53706, USA}
\affiliation{Yale University, New Haven, Connecticut 06520, USA}
\affiliation{LAFEX, Centro Brasileiro de Pesquisas F\'{i}sicas, Rio de Janeiro, Brazil}
\affiliation{Universidade do Estado do Rio de Janeiro, Rio de Janeiro, Brazil}
\affiliation{Universidade Federal do ABC, Santo Andr\'{e}, Brazil}
\affiliation{University of Science and Technology of China, Hefei, People's Republic of China}
\affiliation{Universidad de los Andes, Bogot\'{a}, Colombia}
\affiliation{Charles University, Faculty of Mathematics and Physics, Center for Particle Physics, Prague, Czech Republic}
\affiliation{Czech Technical University in Prague, Prague, Czech Republic}
\affiliation{Institute of Physics, Academy of Sciences of the Czech Republic, Prague, Czech Republic}
\affiliation{Universidad San Francisco de Quito, Quito, Ecuador}
\affiliation{LPC, Universit\'{e} Blaise Pascal, CNRS/IN2P3, Clermont, France}
\affiliation{LPSC, Universit\'{e} Joseph Fourier Grenoble 1, CNRS/IN2P3, Institut National Polytechnique de Grenoble, Grenoble, France}
\affiliation{CPPM, Aix-Marseille Universit\'{e}, CNRS/IN2P3, Marseille, France}
\affiliation{LAL, Universit\'{e} Paris-Sud, CNRS/IN2P3, Orsay, France}
\affiliation{LPNHE, Universit\'{e}s Paris VI and VII, CNRS/IN2P3, Paris, France}
\affiliation{CEA, Irfu, SPP, Saclay, France}
\affiliation{IPHC, Universit\'{e} de Strasbourg, CNRS/IN2P3, Strasbourg, France}
\affiliation{IPNL, Universit\'{e} Lyon 1, CNRS/IN2P3, Villeurbanne, France and Universit\'{e} de Lyon, Lyon, France}
\affiliation{III. Physikalisches Institut A, RWTH Aachen University, Aachen, Germany}
\affiliation{Physikalisches Institut, Universität Freiburg, Freiburg, Germany}
\affiliation{II. Physikalisches Institut, Georg-August-Universität Göttingen, Göttingen, Germany}
\affiliation{Institut f\"{u}r Physik, Universität Mainz, Mainz, Germany}
\affiliation{Ludwig-Maximilians-Universität M\"{u}nchen, M\"{u}nchen, Germany}
\affiliation{Panjab University, Chandigarh, India}
\affiliation{Delhi University, Delhi, India}
\affiliation{Tata Institute of Fundamental Research, Mumbai, India}
\affiliation{University College Dublin, Dublin, Ireland}
\affiliation{Korea Detector Laboratory, Korea University, Seoul, Korea}
\affiliation{CINVESTAV, Mexico City, Mexico}
\affiliation{Nikhef, Science Park, Amsterdam, the Netherlands}
\affiliation{Radboud University Nijmegen, Nijmegen, the Netherlands}
\affiliation{Joint Institute for Nuclear Research, RU-141980 Dubna, Russia}
\affiliation{Institution for Theoretical and Experimental Physics, ITEP, Moscow 117259, Russia}
\affiliation{Moscow State University, Moscow, Russia}
\affiliation{Institute for High Energy Physics, Protvino, Russia}
\affiliation{Petersburg Nuclear Physics Institute, St. Petersburg, Russia}
\affiliation{Instituci\'{o} Catalana de Recerca i Estudis Avançats (ICREA) and Institut de F\'{i}sica d'Altes Energies (IFAE), Barcelona, Spain}
\affiliation{Uppsala University, Uppsala, Sweden}
\affiliation{Lancaster University, Lancaster LA1 4YB, United Kingdom}
\affiliation{Imperial College London, London SW7 2AZ, United Kingdom}
\affiliation{The University of Manchester, Manchester M13 9PL, United Kingdom}
\affiliation{University of Arizona, Tucson, Arizona 85721, USA}
\affiliation{University of California Riverside, Riverside, California 92521, USA}
\affiliation{Florida State University, Tallahassee, Florida 32306, USA}
\affiliation{Fermi National Accelerator Laboratory, Batavia, Illinois 60510, USA}
\affiliation{University of Illinois at Chicago, Chicago, Illinois 60607, USA}
\affiliation{Northern Illinois University, DeKalb, Illinois 60115, USA}
\affiliation{Northwestern University, Evanston, Illinois 60208, USA}
\affiliation{Indiana University, Bloomington, Indiana 47405, USA}
\affiliation{Purdue University Calumet, Hammond, Indiana 46323, USA}
\affiliation{University of Notre Dame, Notre Dame, Indiana 46556, USA}
\affiliation{Iowa State University, Ames, Iowa 50011, USA}
\affiliation{University of Kansas, Lawrence, Kansas 66045, USA}
\affiliation{Louisiana Tech University, Ruston, Louisiana 71272, USA}
\affiliation{Northeastern University, Boston, Massachusetts 02115, USA}
\affiliation{University of Michigan, Ann Arbor, Michigan 48109, USA}
\affiliation{Michigan State University, East Lansing, Michigan 48824, USA}
\affiliation{University of Mississippi, University, Mississippi 38677, USA}
\affiliation{University of Nebraska, Lincoln, Nebraska 68588, USA}
\affiliation{Rutgers University, Piscataway, New Jersey 08855, USA}
\affiliation{Princeton University, Princeton, New Jersey 08544, USA}
\affiliation{State University of New York, Buffalo, New York 14260, USA}
\affiliation{University of Rochester, Rochester, New York 14627, USA}
\affiliation{State University of New York, Stony Brook, New York 11794, USA}
\affiliation{Brookhaven National Laboratory, Upton, New York 11973, USA}
\affiliation{Langston University, Langston, Oklahoma 73050, USA}
\affiliation{University of Oklahoma, Norman, Oklahoma 73019, USA}
\affiliation{Oklahoma State University, Stillwater, Oklahoma 74078, USA}
\affiliation{Brown University, Providence, Rhode Island 02912, USA}
\affiliation{University of Texas, Arlington, Texas 76019, USA}
\affiliation{Southern Methodist University, Dallas, Texas 75275, USA}
\affiliation{Rice University, Houston, Texas 77005, USA}
\affiliation{University of Virginia, Charlottesville, Virginia 22904, USA}
\affiliation{University of Washington, Seattle, Washington 98195, USA}

\author{T.~Aaltonen\ensuremath{^{\dag}}}
\affiliation{Division of High Energy Physics, Department of Physics, University of Helsinki, FIN-00014, Helsinki, Finland; Helsinki Institute of Physics, FIN-00014, Helsinki, Finland}
\author{V.M.~Abazov\ensuremath{^{\ddag}}}
\affiliation{Joint Institute for Nuclear Research, RU-141980 Dubna, Russia}
\author{B.~Abbott\ensuremath{^{\ddag}}}
\affiliation{University of Oklahoma, Norman, Oklahoma 73019, USA}
\author{B.S.~Acharya\ensuremath{^{\ddag}}}
\affiliation{Tata Institute of Fundamental Research, Mumbai, India}
\author{M.~Adams\ensuremath{^{\ddag}}}
\affiliation{University of Illinois at Chicago, Chicago, Illinois 60607, USA}
\author{T.~Adams\ensuremath{^{\ddag}}}
\affiliation{Florida State University, Tallahassee, Florida 32306, USA}
\author{J.P.~Agnew\ensuremath{^{\ddag}}}
\affiliation{The University of Manchester, Manchester M13 9PL, United Kingdom}
\author{G.D.~Alexeev\ensuremath{^{\ddag}}}
\affiliation{Joint Institute for Nuclear Research, RU-141980 Dubna, Russia}
\author{G.~Alkhazov\ensuremath{^{\ddag}}}
\affiliation{Petersburg Nuclear Physics Institute, St. Petersburg, Russia}
\author{A.~Alton\ensuremath{^{\ddag}}\ensuremath{^{jj}}}
\affiliation{University of Michigan, Ann Arbor, Michigan 48109, USA}
% ...

\collaboration{CDF Collaboration}
\altaffiliation[With visitors from]{
\ensuremath{^{a}}University of British Columbia, Vancouver, BC V6T 1Z1, Canada,
\ensuremath{^{b}}Istituto Nazionale di Fisica Nucleare, Sezione di Cagliari, 09042 Monserrato (Cagliari), Italy,
\ensuremath{^{c}}University of California Irvine, Irvine, CA 92697, USA,
\ensuremath{^{d}}Institute of Physics, Academy of Sciences of the Czech Republic, 182~21, Czech Republic,
\ensuremath{^{e}}CERN, CH-1211 Geneva, Switzerland,
\ensuremath{^{f}}Cornell University, Ithaca, NY 14853, USA,
\ensuremath{^{g}}University of Cyprus, Nicosia CY-1678, Cyprus,
\ensuremath{^{h}}Office of Science, U.S. Department of Energy, Washington, DC 20585, USA,
\ensuremath{^{i}}University College Dublin, Dublin 4, Ireland,
\ensuremath{^{j}}ETH, 8092 Z\"{u}rich, Switzerland,
\ensuremath{^{k}}University of Fukui, Fukui City, Fukui Prefecture, Japan 910-0017,
\ensuremath{^{l}}Universidad Iberoamericana, Lomas de Santa Fe, M\'{e}xico, C.P. 01219, Distrito Federal,
\ensuremath{^{m}}University of Iowa, Iowa City, IA 52242, USA,
\ensuremath{^{n}}Kinki University, Higashi-Osaka City, Japan 577-8502,
\ensuremath{^{o}}Kansas State University, Manhattan, KS 66506, USA,
\ensuremath{^{p}}Brookhaven National Laboratory, Upton, NY 11973, USA,
\ensuremath{^{q}}University of Manchester, Manchester M13 9PL, United Kingdom,
\ensuremath{^{r}}Queen Mary, University of London, London, E1 4NS, United Kingdom,
\ensuremath{^{s}}University of Melbourne, Victoria 3010, Australia,
\ensuremath{^{t}}Muons, Inc., Batavia, IL 60510, USA,
\ensuremath{^{u}}Nagasaki Institute of Applied Science, Nagasaki 851-0193, Japan,
\ensuremath{^{v}}National Research Nuclear University, Moscow 115409, Russia,
\ensuremath{^{w}}Northwestern University, Evanston, IL 60208, USA,
\ensuremath{^{x}}University of Notre Dame, Notre Dame, IN 46556, USA,
\ensuremath{^{y}}Universidad de Oviedo, E-33007 Oviedo, Spain,
\ensuremath{^{z}}CNRS-IN2P3, Paris, F-75205 France,
\ensuremath{^{aa}}Universidad Tecnica Federico Santa Maria, 110v Valparaiso, Chile,
\ensuremath{^{bb}}The University of Jordan, Amman 11942, Jordan,
\ensuremath{^{cc}}Universite catholique de Louvain, 1348 Louvain-La-Neuve, Belgium,
\ensuremath{^{dd}}University of Z\"{u}rich, 8006 Z\"{u}rich, Switzerland,
\ensuremath{^{ee}}Massachusetts General Hospital, Boston, MA 02114 USA,
\ensuremath{^{ff}}Harvard Medical School, Boston, MA 02114 USA,
\ensuremath{^{gg}}Hampton University, Hampton, VA 23668, USA,
\ensuremath{^{hh}}Los Alamos National Laboratory, Los Alamos, NM 87544, USA,
\ensuremath{^{ii}}Universit\`{a} degli Studi di Napoli Federico I, I-80138 Napoli, Italy
}
\noaffiliation
\collaboration{D0 Collaboration}
\altaffiliation[With visitors from]{
\ensuremath{^{jj}}Augustana College, Sioux Falls, SD, USA,
\ensuremath{^{kk}}The University of Liverpool, Liverpool, UK,
\ensuremath{^{ll}}DESY, Hamburg, Germany,
\ensuremath{^{mm}}Universidad Michoacana de San Nicolas de Hidalgo, Morelia, Mexico,
\ensuremath{^{nn}}SLAC, Menlo Park, CA, USA,
\ensuremath{^{oo}}University College London, London, UK,
\ensuremath{^{pp}}Centro de Investigacion en Computacion - IPN, Mexico City, Mexico,
\ensuremath{^{qq}}Universidade Estadual Paulista, São Paulo, Brazil,
\ensuremath{^{rr}}Karlsruher Institut f\"{u}r Technologie (KIT) - Steinbuch Centre for Computing (SCC),
\ensuremath{^{ss}}Office of Science, U.S. Department of Energy, Washington, D.C. 20585, USA,
\ensuremath{^{tt}}Thomas Jefferson National Accelerator Facility, Newport News, VA 23606, USA
}
\noaffiliation

\date{\today}

\begin{abstract}
We summarize and combine direct measurements of
the mass of the $W$ boson in $\sqrt{s} = 1.96~\text{TeV}$ proton-antiproton collision  data collected by CDF and D0 experiments at the Fermilab Tevatron Collider. Earlier measurements from CDF and D0 are combined with the two 
latest, more precise measurements: a CDF measurement in
the electron and muon channels using data corresponding to $2.2~ \mathrm{fb}^{-1}$ of integrated
luminosity, and a D0 measurement in the electron channel using data corresponding to $4.3~ \mathrm{fb}^{-1}$ of integrated luminosity.  The resulting Tevatron average for the mass of
the $W$ boson is $\MW = 80\,387 \pm 16~ \text{MeV}$. Including measurements obtained in electron-positron collisions at LEP yields the most precise value of $\MW = 80\,385 \pm 15~ \text{MeV}$. 
\end{abstract}
\pacs{14.70.Fm, 12.15.Ji, 13.38.Be, 13.85.Qk}

\maketitle

\section{Introduction}
\label{intro}

In the standard model (SM), quantum corrections to the mass of the $W$ boson ($\MW$) are dominated by contributions dependent on the mass of the top quark  ($m_t$), the mass of the Higgs boson ($M_H$), and the fine-structure constant $\alpha$.  A precise measurement of $M_W$ and $m_t$ therefore constrains $M_H$. Comparing this constraint with the mass of the Higgs boson recently discovered at the LHC~\cite{lhc-higgs} is a critical test of its nature and the consistency of the SM. Details of the experimental methods used in measurements of $\MW$ are discussed in Ref.~\cite{ashujan}. Prior to the combination reported here, the uncertainty on the world average $\MW$ was 23 MeV~\cite{TEV09,Units}. Direct measurements of $m_t$ at the Fermilab Tevatron collider have a combined uncertainty of 0.94~GeV~\cite{top}, and the uncertainty on $\MW$ would have to be 6 MeV~\cite{cor-rad} to provide  equally  constraining information  on $M_H$. The experimental precision on the measured $\MW$ is therefore currently the limiting factor on the constraints.

The CDF and D0 experiments at the Fermilab Tevatron proton-antiproton collider reported several direct measurements of the natural width~\cite{Wwidth} and mass~\cite{MW-CDF-RunZ,MW-CDF-RUN1A,MW-CDF-RUN1B,MW-CDF-RUN2a,MW-D0-Ia,MW-D0-I,MW-D0-I-rap,MW-D0-I-edge,MW-D0-RUN2a,MW-CDF-RUN2,MW-D0-Run2b} of the $W$ boson, using the $e\nu_e$ and $\mu\nu_\mu$ decay modes of the $W$ boson.  Measurements of $\MW$ have been reported by CDF with data sets collected during 1988-1989~\cite{MW-CDF-RunZ}, 1992-1993~\cite{MW-CDF-RUN1A}, 1994-1995~\cite{MW-CDF-RUN1B}, and 2001-2004~\cite{MW-CDF-RUN2a} and by D0 using data taken during 1992-1995~\cite{MW-D0-Ia,MW-D0-I,MW-D0-I-rap,MW-D0-I-edge} and 2002-2006~\cite{MW-D0-RUN2a}. 

This article describes a combination of $\MW$ measurements including recent measurements from CDF using the 2002-2007 dataset~\cite{MW-CDF-RUN2} and D0 using the 2006-2009 dataset~\cite{MW-D0-Run2b} denoted below as CDF (2012) and D0 (2012), respectively. The recent CDF (2012) measurement supersedes the previous measurement~\cite{MW-CDF-RUN2a}, which was based on an integrated luminosity of $200\ \mathrm{pb}^{-1}$ and was used in previous combinations~\cite{TEV08, TEV09}. The combination takes into account the statistical and systematic uncertainties as well as correlations among systematic uncertainties and supersedes the previous combinations~\cite{MWGW-RunI-PRD,TEV08, TEV09}.  All the combinations presented in this article are done using the best linear unbiased estimator (BLUE) method~\cite{BLUE}, which prescribes the construction of a covariance matrix from partially correlated measurements.

\section {$\bm{W}$-Boson Mass Measurement Strategy at the Tevatron}
\label{strategy}

At the Tevatron, $W$ bosons are primarily produced in quark-antiquark annihilation, $q\overline{q'}$ $\rightarrow$ $W$ + $X$, where $X$ can include QCD radiation, such as initial-state gluon radiation, that results in measurable hadronic-recoil energy.  The $W$-boson mass is measured using low-background samples of $W\rightarrow \ell\nu_\ell$ decays ($\ell=e,\mu$ at CDF and $\ell=e$ at D0) that are reconstructed using the CDF~\cite{cdf_det} and D0~\cite{d0_det} detectors. 
The mass is determined using three kinematic variables measured in the plane perpendicular to the beam direction: the transverse momentum of the charged lepton ($p_{T}^{\ell}$), the transverse momentum of the neutrino ($p_{T}^{\nu}$), and the transverse mass $m_{T}^{\ell} = \sqrt {2p_{T}^{\ell}p_{T}^{\nu}(1 - \cos\Delta\phi)}$, where $\Delta\phi$ is the opening angle between the lepton and neutrino momenta in the plane transverse to the beam. The magnitude and direction of $p_{T}^{\nu}$ is inferred from the vector of the missing transverse energy $\METl$~\cite{Etmiss}. 
The $W$-boson mass is extracted from maximum-likelihood fits to the binned distributions of the observed $p_T^{\ell}$, $\METl$, and $m_{T}^{\ell}$ values using a parametrized  simulation of these distributions as a function of $\MW$. These simulations depend on the kinematic distributions of the $W$-boson decay products and also on detector effects that are constrained using theoretical calculations and control samples. The kinematic distributions are determined by several effects including the $W$-boson transverse momentum $p_T(W)$ and the parton distribution functions (PDFs) of the interacting protons  and antiprotons. Major detector effects include energy response to leptons, hadronic recoil, the response to QED radiation, and multiple-interaction pileup, together with calorimeter acceptance effects and lepton-identification efficiencies.  The detailed simulations developed at CDF and D0 enable the study of these effects to better than 1 part in $10^4$ precision on the observed value of $\MW$.

In the CDF (2012)  and D0 (2012) measurements, the kinematic properties of $W$-boson production and decay are simulated using {\sc resbos}~\cite{resbos}, which is a next-to-leading order generator that includes next-to-next-to-leading logarithm resummation of soft gluons at low boson $p_T$~\cite{blny}. The momenta of interacting partons in {\sc resbos} are calculated as a fractions of the colliding (anti)proton momenta using the CTEQ6.6 \cite{cteq66} PDFs. The radiation of photons from final-state leptons is simulated using {\sc photos}~\cite{ref:PHOTOS}.

\section{CDF (2012) and D0 (2012)  Measurements}
\label{newmeas}

\subsection{CDF Measurement}
\label{cdf}

The CDF (2012) measurement uses data corresponding to an integrated luminosity of $2.2~ \mathrm{fb}^{-1}$, collected between 2002 and 2007. Both the muon ($W\to\mu\nu_\mu$) and electron ($W\to e\nu_e$) channels are considered.  Decays of $J/\psi$ and $\Upsilon$ mesons into muon pairs are reconstructed in a central tracking system to establish the absolute momentum scale. A measurement of the $Z$-boson mass ($\MZ$) in $Z\to\mu\mu$ decays is performed as a consistency check. This measurement, which uses the tracking detector, yields $M_Z = 91\,180 \pm 12 ~\text{(\rm stat)} \pm 10 ~\text{(\rm syst)}~\text{MeV}$, consistent with the world average mass of $91\,188 \pm 2~\text{MeV}$~\cite{lep}, and is therefore also used as an additional constraint on the momentum scale. The electromagnetic calorimeter energy scale and nonlinearity are determined by fitting the peak of the $E/p$ distribution of electrons from $W$ $\rightarrow$ $e \nu$ and $Z$ $\rightarrow$ $ee$ decays, where $E$ is the energy measured in the calorimeter and $p$ is the momentum of the associated charged particle. The lower tail of the $E$/$p$ distribution is used to determine the amount of material in the tracking detector. The $Z$-boson mass measured in $Z\rightarrow ee$ decays is used as a consistency check and to constrain the energy scale. The value of $M_Z = 91\,230 \pm 30 ~\text{(\rm stat)}  \pm 14 ~\text{(\rm syst)}~\text{MeV}$ from the calorimetric measurement is also consistent with the world average. 

The CDF (2012) measurement of $\MW$ is obtained from the combination of six observables: $\ptmu$, $\METmu$, $\mtmu$, $\pte$, $\METe$ and $\mte$. The combined result is $\MW$ = $80\,387 \pm 12 ~\text{(\rm stat)} \pm 15 ~\text{(\rm syst)}~ \text{MeV}$.   
 Table \ref{cdferr} summarizes the sources of  uncertainty in the CDF measurement.

\begin{table}
\begin{ruledtabular}
\begin{tabular}{lr}
Source & Uncertainty (MeV)   \\ 
\hline 
Lepton energy scale and resolution             & \syspscalecombin \ \\
Recoil energy scale and resolution            &   6  \\
Lepton removal from recoil           &    \sysholecombin \    \\
Backgrounds               &    \sysbkgcombin\       \\
\hline
   Experimental subtotal  & 10 \\
\hline
Parton distribution functions      &   \syspdfcombin \ \\
QED radiation             &  \sysqedcombin\    \\
$p_T(W)$ model  &  \sysgtwocombin \  \\ 
\hline
Production subtotal &12\\ 
\hline
Total systematic uncertainty                     & 15\\

$W$-boson event yield     &    \statcombin\     \\
\hline
Total  uncertainty                   &    \bluealle\     \\

\end{tabular}
\caption{Uncertainties of the CDF  (2012) $\MW$ measurement determined from the combination of the  six measurements.}
\label{cdferr}
\end{ruledtabular}
\end{table}

\subsection{D0 Measurement}
\label{d0}
The D0 (2012) measurement uses data corresponding to $4.3$ fb$^{-1}$ of integrated luminosity recorded between 2006 and 2009. D0 calibrates the calorimeter energy scale using $Z$ $\rightarrow$ $ee$ decays. Corrections for energy lost in un-instrumented regions are based on a comparison between the shower-development profiles from data and from a detailed {\sc geant}-based simulation~\cite{geant} of the D0 detector. The world average value for $M_Z$~\cite{lep} is used to determine the absolute energy-scale of the calorimeter, which is thereafter used to correct the measurement of the electron energy from the $W$-boson decay. This $\MW$ measurement is therefore equivalent to a measurement of the ratio of $W$- and $Z$-boson masses.  This calibration method eliminates many systematic uncertainties common to the $W$- and $Z$-boson mass measurements, but its precision is limited by the size of the available $Z$-boson data set. 

 \begin{table}
 \begin{ruledtabular}
\begin{center}

\begin{tabular}{lr}

Source & Uncertainty (MeV)   \\

  \hline 
  Electron energy calibration       & 16  \\
  Electron resolution model        &  2  \\
  Electron shower modeling          &  4  \\
  Electron energy loss model     &  4  \\
  Recoil energy scale and resolution     & 5  \\
  Electron efficiencies      &  2  \\
  Backgrounds                  &  2  \\ \hline
    Experimental subtotal          & 18 \\ 
\hline				    				     
  Parton distribution functions              & 11  \\
  QED  radiation               &  7  \\
  $p_T(W)$ model          &  2  \\ \hline
  Production subtotal       &  13\\ 
  \hline
  Total systematic uncertainty                     & 22 \\

  $W$-boson event yield & 13\\
  \hline 
  Total uncertainty &26 \\
    \end{tabular}
\end{center} 
\caption{Uncertainties  of the D0  (2012) $\MW$  measurement determined from the combination of   the two most sensitive observables  $\mte$ and $\pte$.}
\label{table:d0err}
\end{ruledtabular}
\end{table}

 The results obtained with the two most sensitive observables $\mte$ and $\pte$ are combined to determine the $W$-boson mass of $\MW = 80\,367 \pm 13 ~\text{(stat)} \pm 22 ~\text{(syst)}~\text{MeV}$. A summary of the uncertainties is presented in Table~\ref{table:d0err}. 
This D0 (2012) measurement is combined with a previous D0 measurement~\cite{MW-D0-RUN2a} corresponding to an integrated luminosity of 1.0 fb$^{-1}$, which uses data recorded between 2002 and 2006, to yield $\MW = 80\,375 \pm 11 ~\text{(stat)} \pm 20 ~\text{(syst)}~\text{MeV}$.

\section {Combination with previous Tevatron Measurements}
\label{comb}

The CDF measurements from Ref.~\cite{MW-CDF-RunZ} (1988-1989) and Ref.~\cite{MW-CDF-RUN1A} (1992-1993) were made using superseded PDF sets and have been corrected~\cite{TEV08} using recent PDF sets.  The previous results are  also adjusted to use the same combination technique (the BLUE method) as in later combinations. The  templates for fitting $\MW$  assume the Breit-Wigner running-width scheme propagator, $1/{(\hat{s}-\MW^2+i\hat{s}\GW/\MW)}$, which makes the value of $\MW$ determined by the fit dependent on $\GW$. Here, $\hat{s}$ is the square of the center-of-mass energy in the parton reference frame and $\GW$ is the total width of the $W$ boson. Different measurements have used different values of $\GW$, yielding a shift in measured values of the $W$-boson mass~\cite{TEV08}, $\Delta M_W=-(0.15\pm0.05)~\Delta\GW$, where $\Delta\GW$ is the difference  between the value of $\GW$ predicted by the SM, $\GW = 2\,092.2 \pm 1.5~\text{MeV}$~\cite{PR}, and that used in a particular analysis. The prediction of $\GW$ assumes $\MW = 80\,385 \pm 15~\text{MeV}$, which is a preliminary world-average combination result~\cite{cdf-d0-wmass2012} of this article. The impact of the corrections on the final $\MW$ combination reported in this article is found to be less than $0.2$~MeV. Table \ref{tab:MW-inputs12} summarizes all inputs to the combination and the corrections made to ensure consistency across measurements.

\begin{table*}
\begin{ruledtabular}
\begin{tabular}{l c c c c c c c c}

  & CDF~\cite{MW-CDF-RunZ} & CDF~\cite{MW-CDF-RUN1A} & CDF~\cite{MW-CDF-RUN1B} &
   D0~\cite{MW-D0-Ia,MW-D0-I,MW-D0-I-rap,MW-D0-I-edge} & D0~\cite{MW-D0-RUN2a} &CDF~\cite{MW-CDF-RUN2} & D0~\cite{MW-D0-Run2b} \\
  & {\small 4.4~pb$^{-1}$}  & {\small 18.2~pb$^{-1}$} &   {\small 84~pb$^{-1}$}  & {\small 95~pb$^{-1}$}  & {\small 1.0 fb$^{-1}$} & \small{2.2 fb$^{-1}$} & \small{4.3 fb$^{-1}$} \\
&{\small (1988-1989)} & {\small (1992-1993)} &{\small (1994-1995)} &{\small (1992-1995)} &{\small (2002-2006)} &{\small  (2002-2007)} &{\small (2006-2009)}  \\  
\hline

 Mass and width  &  &  &  &  &  &  &  \\    
 $\MW$  & 79\,910 & 80\,410 & 80\,470 & 80\,483 & 80\,400 & 80\,387 & 80\,367 \\
  \text{$\Gamma_W$ } & 2\,100 & 2\,064 & 2\,096 & 2\,062 & 2\,099 & 2\,094 & 2\,100 \\
  \hline
 $\MW$ uncertainties &&&&&&&\\
 \text{PDF} & 60 & 50 & 15 & 8 & 10& 10 & 11 \\
 \text{Radiative  corrections} & 10 & 20 & 5 & 12 & 7 & 4 & 7 \\
 \text{$\Gamma_W$} & 0.5 & 1.4 & 0.3 & 1.5 & 0.4 & 0.2 & 0.5\\
  \text{Total} & 390 & 181& 89 & 84 & 43 & 19 & 26 \\
 \hline
  $\MW$ corrections &&&&&&&\\
   \text{$\Delta \GW$} & +1.2 & $-4.2$ & +0.6 & $-4.5$ & +1.1 & +0.3 & +1.2 \\
 \text{PDF} & +20 & $-25$ & 0 & 0 & 0 & 0 & 0 \\
 \text{Fit method} & $-3.5$ & $-3.5$ & $-0.1$ & 0 & 0 & 0 & 0 \\
 \text{Total} & +17.7 & $-32.7$ & +0.5 & $-4.5$ & +1.1 & +0.3 & +1.2 \\
 \hline
 $\MW$ corrected & 79\,927.7 & 80\,377.3 & 80\,470.5 & 80\,478.5 & 80\,401.8 & 80\,387.3 & 80\,368.6 \\
\end{tabular}
\caption{The input data used in the $\MW$ combination. All entries are in units of MeV.
\label{tab:MW-inputs12}}
\end{ruledtabular}
\end{table*}

\section{Correlations  in the CDF and D0  $\bm{M_W}$  measurements}
\label{correl}

The increased statistical power of CDF (2012) and D0 (2012) $\MW$ measurements necessitates a more detailed treatment of the systematic uncertainties due to the $W$-boson production and decay model that are independent of the data-sample size. We assume that for each uncertainty category, the smallest uncertainty across measurements  is fully correlated  while excesses above that level are generally assumed to be due to uncorrelated differences between measurements. One exception corresponds to the two D0 measurements that use very similar models and are treated as fully correlated~\cite{MW-D0-RUN2a,MW-D0-Run2b}.

The experimental systematic uncertainties of the D0 measurement are dominated by the
uncertainty in the energy scale for electrons and are nearly purely of statistical origin, as they are  derived from the limited sample of $Z\to ee$ decays.  CDF uses independent data from the central tracker to set the muon and electron energy scales. Thus, we assume no correlations between the experimental uncertainties of CDF and D0, or between independent measurements by either experiment.

Three sources of systematic uncertainty due to modeling of the production and decay of $W$ and $Z$ bosons are assumed to be at least partially correlated across all Tevatron measurements: (1) the choice of PDF sets, (2) the assumed $\GW$ value, and (3) the electroweak radiative corrections. 
 
 \subsection {PDF sets}
Both experiments use the CTEQ6.6~\cite{cteq66} PDF set in their $W$-boson production model. D0 uses the CTEQ6.1~\cite{CTEQ61M} uncertainty set to estimate the PDF uncertainties, while CDF uses MSTW2008~\cite{mstw2008} and checks consistency with the CTEQ6.6 uncertainty set. Since these PDF sets are similar and rely on common inputs, the uncertainties introduced by PDFs in the recent measurements are assumed to be correlated and treated using the prescription for partial correlations described above.

\subsection {Assumed $\GW$ value} 
We assume that the small uncertainty due to $\GW$ is fully correlated across all measurements.

\subsection {QED radiative corrections}
Current estimates of the uncertainties due to electroweak radiative corrections include a significant statistical component due to the size of the simulated data sets used in the uncertainty-propagation studies.  The {\sc photos}~\cite{ref:PHOTOS} radiative correction model is used  in the recent measurements with consistency checks from {\sc w(z)grad}~\cite{ref:WGRAD} and {\sc horace}~\cite{horace}. These studies yield model differences consistent within statistical uncertainties. We assume that uncertainties from purely theoretical sources, totaling 3.5 MeV, are correlated while remaining uncertainties, partially dependent on detector geometry, are uncorrelated. 

\section{Combination of Tevatron $\bm{M_W}$ measurements}
\label{cov}
The measurements of $\MW$ obtained at Tevatron experiments included in this combination are given in Table~\ref{tab:MW-inputs12} and include both the latest measurements~\cite{MW-CDF-RUN2, MW-D0-Run2b} discussed above, but exclude the superseded 0.2~fb$^{-1}$~CDF measurement~\cite{MW-CDF-RUN2a}.  Table \ref{contribution} shows the relative weight of each measurement in the combination. The combined value of the $W$-boson mass obtained from measurements performed at Tevatron experiments is

\begin{equation}
M_{W} = 80\,387 \pm 16~\textrm{MeV}.
\end{equation}

The $\chi^{2}$ for the combination is 4.2 for 6 degrees of freedom, with a probability of 64\%. The global correlation matrix for the seven measurements is shown in Table~\ref{global}.

\begin{table}
\begin{ruledtabular}
\begin{center}
\begin{tabular}{l r}   
Measurement           &    Relative weight in \%  \\ \hline
CDF~\cite{MW-CDF-RunZ}      &   0.1  \\
CDF~\cite{MW-CDF-RUN1A}      &   0.5  \\
CDF~\cite{MW-CDF-RUN1B}     &   1.9  \\
D0~\cite{MW-D0-Ia,MW-D0-I,MW-D0-I-rap,MW-D0-I-edge}     &  2.8  \\
D0~\cite{MW-D0-RUN2a}    & 7.9 \\
CDF~\cite{MW-CDF-RUN2}    &  60.3  \\ 
D0~\cite{MW-D0-Run2b}     &  26.5  
\end{tabular}
\caption{Relative weights of the contributions to the combined Tevatron measurement of $\MW$. \label{contribution}}
\end{center}
\end{ruledtabular}
\end{table}

\begin{table*}
\begin{ruledtabular}
\begin{center}
\small{
\begin{tabular}{ l  l  l  l  l  l  l  l  l  }    
& CDF~\cite{MW-CDF-RunZ} & CDF~\cite{MW-CDF-RUN1A} & CDF~\cite{MW-CDF-RUN1B} &
   D0~\cite{MW-D0-Ia,MW-D0-I,MW-D0-I-rap,MW-D0-I-edge} & D0~\cite{MW-D0-RUN2a} & CDF~\cite{MW-CDF-RUN2} &D0~\cite{MW-D0-Run2b} \\
\hline  
CDF~\cite{MW-CDF-RunZ}   						     &    1 & 0.002 & 0.003 & 0.002 & 0.007 & 0.015 & 0.011 \\
CDF~\cite{MW-CDF-RUN1A} 							 &      &     1 & 0.007 & 0.005 & 0.014 & 0.033 & 0.024 \\
CDF~\cite{MW-CDF-RUN1B}  							 &      &       &     1 & 0.009 & 0.029 & 0.066 & 0.049 \\
D0~\cite{MW-D0-Ia,MW-D0-I,MW-D0-I-rap,MW-D0-I-edge}  &      &       &       &     1 & 0.019 & 0.044 & 0.032 \\
D0~\cite{MW-D0-RUN2a}							     &	    &	    &	    &	    & 	  1 & 0.137 & 0.137 \\
CDF~\cite{MW-CDF-RUN2} 							     &      &       &       &       &       &     1 & 0.230 \\
D0~\cite{MW-D0-Run2b}                                &      &       &       &       &       &       & 1\\ 
\end{tabular}
}
\caption{Correlation coefficients among measurements. \label{global}}
\end{center}
\end{ruledtabular}
\end{table*}

\section{World Average}
\label{world}
\begin{figure}
\includegraphics [width=0.48\textwidth] {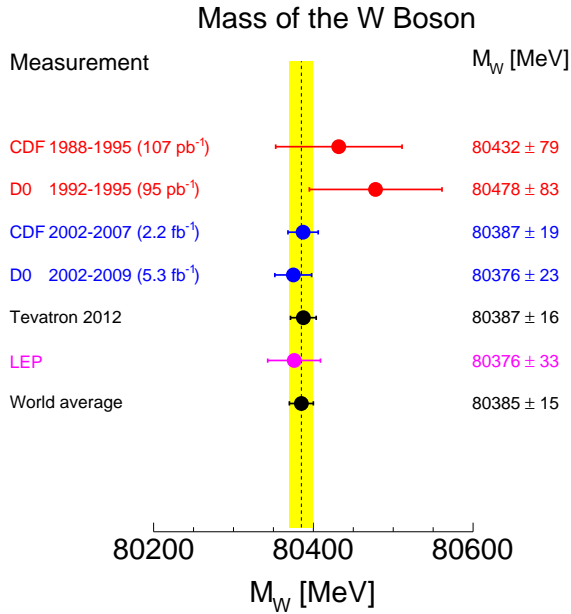}
\caption{$W$-boson mass determinations from the CDF and D0 Run I (1989 to 1996) and Run II (2001 to 2009) measurements, the new Tevatron average, the LEP combined result ~\cite{lep}, and the world average obtained by combining the Tevatron and LEP averages assuming no correlations between them. The world-average uncertainty (15 MeV) is indicated by the shaded band.}
\label{fig:sum}
\end{figure}

We also combine the Tevatron measurements with the value $\MW = 80\,376 \pm 33~\text{MeV}$  determined from $e^{+}\ e^{-} \rightarrow W^+W^-$ production at LEP~\cite{lep}.
Assuming no correlations, this yields the currently most precise value of the $W$ boson mass~of

\begin{equation}  
\MW = 80\,385 \pm 15~\textrm{MeV}.  
\end{equation}
The combination of the seven statistically independent Tevatron measurements and the LEP measurement yields  a $\chi^2$ of 4.3 for 7 degrees of freedom with a probability of 74\%. Figure~\ref{fig:sum} shows the individual measurements and the most recent combined world average of $\MW$.

\section{Summary}
\label{sum}

\begin{figure}
\includegraphics [scale=0.45] {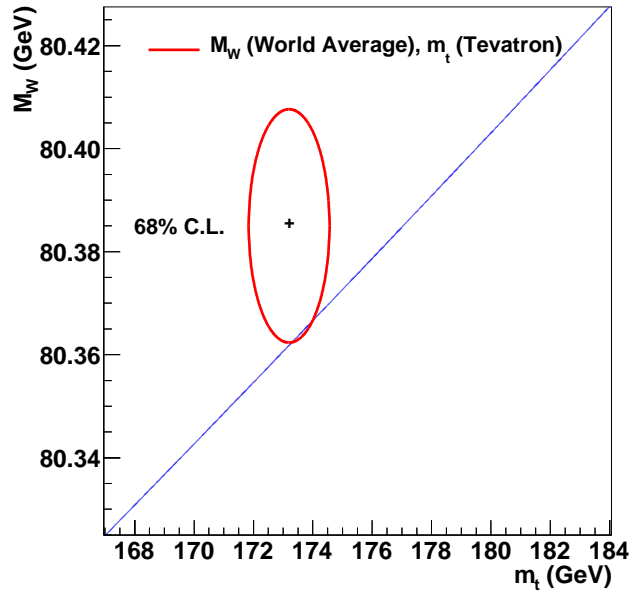}
\caption{
 The most recent world average of $M_W$ is displayed along with the mass of the top quark $m_t$~\cite{top} at 68\% C.L. by area. The diagonal line is the indirect prediction of $M_W$ as a function of $m_t$, in the SM given by Ref.~\cite{cor-rad}, assuming the measurements of the ATLAS and CMS~\cite{lhc-higgs} experiments of the candidate Higgs-boson masses of 126.0 GeV and 125.3 GeV respectively.} 
\label{fig:contour}
\end{figure}

The latest high-precision measurements of $\MW$ performed at the CDF and D0 experiments, combined with previous measurements by the Tevatron experiments, improve the uncertainty on the combined  Tevatron $\MW$ value to 16~MeV. The combination of this measurement with the LEP average for $\MW$ further reduces the uncertainty to 15~MeV. The substantial improvement in the experimental precision on $M_W$ leads to tightened indirect constraints on the mass of the SM Higgs boson. The direct measurements of the mass of the Higgs boson at the LHC~\cite{lhc-higgs} agree, at the level of  1.3 standard deviations, with these tightened indirect constraints~\cite{gfitter}. This remarkable success of the standard model is also shown in Fig.~\ref{fig:contour}, which includes the new world average $W$-boson mass, the Tevatron average top-quark mass measurement~\cite{top}, and shows consistency among these with the calculation of $\MW$~\cite{cor-rad}, assuming Higgs-boson mass determinations from the ATLAS and CMS experiments~\cite{lhc-higgs}. 

%
We thank the Fermilab staff and the technical staffs of the participating institutions for their vital contributions
and acknowledge support from the
DOE and NSF (USA);
ARC (Australia);
CNPq, FAPERJ, FAPESP, and FUNDUNESP (Brazil);
NSERC (Canada);
CAS and CNSF (China);
Colciencias (Colombia);
MSMT and GACR (Czech Republic);
the Academy of Finland;
CEA and CNRS/IN2P3 (France);
BMBF and DFG (Germany);
DAE and DST (India);
SFI (Ireland);
INFN (Italy);
MEXT (Japan);
the Korean World Class University Program and NRF (Korea);
CONACyT (Mexico);
FOM (The Netherlands);
MON, NRC KI, and RFBR (Russia);
the Slovak R\&D Agency (Slovakia);
the Ministerio de Ciencia e Innovaci\'{o}n, and Programa Consolider-Ingenio 2010 (Spain);
the Swedish Research Council (Sweden);
SNSF (Switzerland);
STFC and the Royal Society (United Kingdom);
and the A. P. Sloan Foundation (USA).


\label{sec:thebib}


\begin{thebibliography}{99}
\bibitem{Units}
We use units where $c=1$ throughout this article.
\bibitem{Etmiss}
$\METl = -(\vec{p}_T^{\ell} + \vec{u}_T)$ where $\vec{u}_T$ is the vector sum of calorimetric $\vec{p}_T$ excluding the charged lepton(s).
\end{thebibliography}
\end{document}